\documentclass[aps,prb,twocolumn,showpacs,superscriptaddress]{revtex4-1}
\usepackage{graphicx}
\usepackage{dcolumn}
\usepackage{bm}
\usepackage{color}%
\usepackage{amsmath}
\usepackage{amsfonts}

\begin{document}

\title{Decay behavior of localized states at reconstructed armchair graphene edges}

\author{Changwon Park}
\affiliation{Department of Physics and Astronomy, Seoul National University, Seoul 151-747, Korea}
\affiliation{Center for Nanophase Materials Sciences, Oak Ridge National Laboratory, Oak Ridge, Tennessee 37831, United States}

\author{Jisoon Ihm}
\affiliation{Department of Physics and Astronomy, Seoul National University, Seoul 151-747, Korea}

\author{Gunn Kim}
\email[The author to whom correspondence should be addressed: ]{gunnkim@sejong.ac.kr}
\affiliation{Department of Physics and Graphene Research Institute, Sejong University, Seoul 143-747, Korea}

\begin{abstract}

Density functional theory calculations are used to investigate the
electronic structures of localized states at reconstructed armchair
graphene edges. We consider graphene nanoribbons with two different
edge types and obtain the energy band structures and charge
densities of the edge states. By examining the imaginary part of the
wavevector in the forbidden energy region, we reveal the decay
behavior of the wavefunctions in graphene. The complex band
structures of graphene in the armchair and zigzag directions are
presented in both tight-binding and first-principles frameworks.
\end{abstract}


\maketitle

\section{Introduction}

One-dimensional boundaries are a distinctive feature of finite-sized
graphene, which means that an investigation and understanding of the
electronic properties of graphene edges are of particular
importance. Even before the first production of graphene
from graphite by mechanical exfoliation in 2004\cite{graphene}, the edge state
was shown to be a typical example of the manifestation
of the topological characteristics of a bulk band\cite{topology}.
The boundary or edge effects become more significant for practical
applications as the size of graphene in a device becomes smaller.
Real graphene edges are usually passivated with functional groups,
transition metals, or hydrogen atoms depending on the specific purpose,
but even without these external chemicals or elements,
reconstruction of the graphene edge itself has been observed\cite{Koskinen}.
How deeply the edge-induced state penetrates the bulk determines
the decay length and extent of the state localization.

According to Bloch's theorem, the eigenstates of the
single-electron Schr\"odinger equation in a crystal satisfy
$\psi({\bf r})=e^{i{\bf k}\cdot {\bf r}}u_{n {\bf k}}({\bf r}),$
where $u_{n {\bf k} }$ is a function that has the same periodicity
as the crystal, $n$ is the band index, and {\bf k} is the
wavevector. In the case of an infinite crystal, the Born--von Karman
cyclic boundary conditions\cite{Ashcroft} restrict the wavevectors
to real quantities. However, complex Bloch {\bf k} vectors are
allowed for finite crystals, and the complex band structure of a
periodic system is the conventional band structure extended to
complex Bloch {\bf k} vectors. Near a crystal surface or interface,
one can match a wavefunction with a complex {\bf k} between the
inside and outside of the crystal region, and thus surface or
interface evanescent states arise\cite{Kohn, Heine1, Heine2, Chang}.

The complex band structure concept can also be adopted for graphene
edges; the properties of the edge states are closely related to the
band structure of infinite graphene.
If we know the dispersion relation of the edge state that can be
accurately calculated in a relatively narrow graphene nanoribbon
(GNR), then this approach allows us to predict the decay behavior of
the edge state in semi-infinite graphene, combined with the complex
band structure of infinite (bulk) graphene.
The wavevector can be split into a component parallel to the edge, $k_{||}$, which is conserved
during scattering, and a perpendicular component, $k_{\perp}$. Then,
for each real $k_{||}$, the dispersion relation $E = E(k_{\perp})$
allows a complex $k_{\perp} = k' + i\kappa$, where $k'$ and $\kappa$
are real. We refer to the imaginary part $\kappa$ as the decay
parameter.

In this paper, we focus on reconstructed armchair graphene edges. A
perfect armchair graphene edge has no localized edge state since its
one-dimensional bulk Hamiltonian $H_{k_{||}}(k_{\perp})$ is
topologically trivial for any $k_{\|}$ value\cite{topology}.
However, reconstructed armchair edges may have localized edge states because
of the modification of the geometries and hopping properties. Using
first-principles calculations, we show that the possible localized
edge states have decay behaviors that are associated with the
complex band structure of graphene in the bulk.
{Near a graphene edge, the solution for an edge state should be
matched to the bulk graphene wavefunction with complex {\bf k}.
Therefore, to investigate the decay pattern of the localized edge
states, we need to calculate the complex band structure of
graphene.}
In addition, we provide analytic solutions to the complex band structures for both
the armchair and zigzag directions using the nearest-neighbor
tight-binding method.

\begin{figure}[t]
\includegraphics[width=1.0\columnwidth]{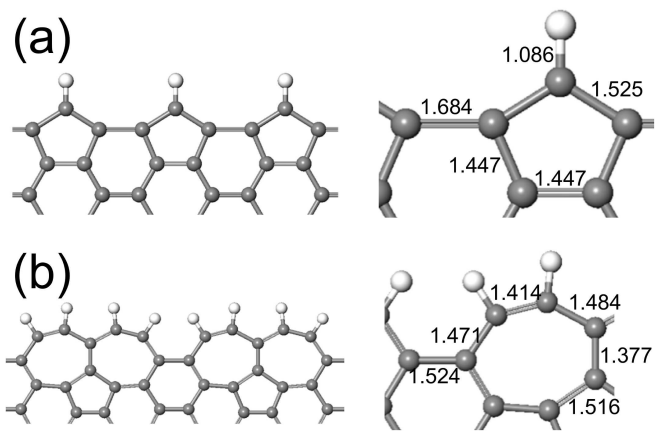}
\caption{(Color online) Model structures of reconstructed armchair edges: (a)
pentagonal armchair edge with missing carbon atoms [ac(56)], and (b)
reconstructed armchair edge in which two heptagons created by the
Stone--Wales transformation share a side [ac(677)]. The bond lengths
(in \AA) at the graphene edges are shown in the right panel. The
gray and white spheres are carbon and hydrogen atoms, respectively.}

\label{Fig1}
\end{figure}

\section{Computational details}

We performed density functional theory calculations within the
generalized gradient approximation (GGA) for the
exchange-correlation functional. The Perdew--Burke--Ernzerhof (PBE)
functional form\cite{PBE} was adopted for the GGA. Ionic potentials
were described by the projector augmented wave (PAW)\cite{PAW}
method implemented in the Vienna Ab Initio Simulation Package
(VASP)\cite{VASP}.
To mimic semi-infinite graphene, we chose GNRs that have an ac(56)-
or ac(677)-type edge\cite{Koskinen} on one side and a perfect
armchair edge on the other side. The widths of our model GNRs with
the ac(56)- and ac(677)-type edges were $\sim$5 and $\sim$6 nm,
respectively.
We used 24 and 12 {\bf k} points to sample the
Brillouin zone (BZ) in the edge direction in the respective edge
geometries. A plane-wave energy cutoff of 400 eV was used for the
structural relaxation, which continued until the atomic forces were
smaller than 20 meV/\AA. The size of the unit cell of the
ac(677)-type edge was twice that of the ac(56)-type edge along the
ribbon axis, as shown in Fig.~\ref{Fig1}. The supercells contained
91 and 204 carbon atoms for the GNRs with the ac(56)- and
ac(677)-type edges, respectively.

To calculate the complex band structures in the primitive cell, we
employed the Quantum Espresso package\cite{pwscf}. The Vanderbilt
ultrasoft pseuopotential\cite{Vanderbilt} was generated through the
Rappe--Rabe--Kaxiras--Joannopoulos (RRKJ)\cite{RRKJ}
pseudo-wavefunction construction scheme. The kinetic energy cutoff
was 30 Ry. From the symmetry of the hexagonal cell, the net force
exerted on each carbon atom was zero, and the lattice parameters
were optimized by the Birch--Murnaghan equation of
state\cite{Murnaghan, Birch}.
To consider the conservation of ${\bf k}_{||}$ in the presence of
the edge and the crystallographic direction to the edge, we
considered a rectangular unit cell containing 4 carbon atoms for the
armchair and zigzag edges. This is the smallest unit cell with the lattice vector aligned along ${\bf
k}_{||}$.
All results are obtained in the rectangular unit cell in the present study.

For the {\bf k} point sampling, 24
$\times$ 24 $\times$ 1 {\bf k} points were chosen in the
Monkhorst--Pack scheme\cite{Monkhorst-Pack}.

In the tight-binding approximation, the Hamiltonian of a graphene
monolayer is expressed as
\begin{equation}
H= \sum_{<i,j>} \left (t c_i^{\dagger}c_j + {\rm h.c.}\right ),
\end{equation}
where $\sum_{<i,j>}$ sums over only nearest-neighbor pairs,
$c_i^{\dagger}$($c_i$) is the creation (annihilation) operator, and
h.c. indicates the Hermitian conjugate. Here, $t$ is the
nearest-neighbor hopping energy ($= -2.88$ eV).

\begin{figure}[t]
\includegraphics[width=0.8\columnwidth]{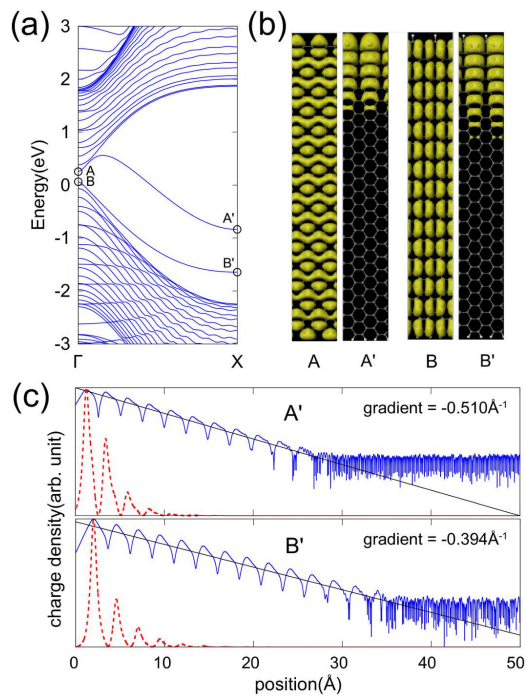}
\caption{(Color online) (a) Band structure of a GNR with an ac(56) edge, and (b)
electronic densities of two edge-related states. The labels A (B)
and A$'$ (B$'$) indicate electronic densities in the same band at
$\Gamma$ and X, respectively. (c) Plane-averaged electron densities
(red) and their semilog plots (blue) of the GNR with the ac(56)
edge showing an exponentially decaying character; the gradients of
the fitted curves give the decay parameters in units of \AA$^{-1}$.
The position indicated on the horizontal axis is the distance from
the hydrogen atom
at the ac(56) edge to the middle of the ribbon.}
\label{Fig2}
\end{figure}

\section{Results and discussion}

The optimized structures of the two GNRs are shown in
Fig.~\ref{Fig1}. The ac(56) model shows a pentagonal reconstruction
of the armchair edge (with a connecting hexagon) that requires the
diffusion of carbon atoms\cite{Koskinen}. The ac(677)
model\cite{Koskinen, GDLee}, however, shows a reconstruction in
which two heptagons share a side (a carbon bond). Unlike the models
of Koskinen {\it et al}.\cite{Koskinen}, our graphene edges are
passivated with hydrogen atoms.

Figure \ref{Fig2} shows the electronic band structure and charge
densities at $\Gamma$ and X in the armchair GNR with the pentagonal
reconstruction, ac(56). Two bands labeled as AA$'$ and BB$'$ are
seen in the forbidden energy band. Because of the added carbon
atoms, one additional band, labeled as AA$'$, appears near the Fermi
level. In terms of the tight-binding calculation, this corresponds
to the inclusion of new basis functions. At the zone boundary (X),
the edge states exist as deep levels in the forbidden energy region
so that they are more localized than the edge states at the $\Gamma$
point. The overlapping of a localized state with the bulk states in
{\bf k}-space is generally referred to as {\em surface
resonance}\cite{Kolasinski}; the localized state penetrates the bulk
and couples strongly to the bulk states. Thus, the electronic states
labeled A and B for the ac(56) model are extended GNR states with
enhanced amplitudes near the edge.

In our previous study\cite{PNAS} of the reconstructed armchair edge
ac(56), we showed that the edge hopping energy $t_0$ at the pentagon
is approximately $-2$ eV by using the maximally localized Wannier
function method\cite{Marzari}. In ideal graphene, the hopping energy
$t$ is $-2.88$ eV, which thus corresponds to boundary softening
($t_0 < t$). Li {\em et al.}\cite{WLi} showed that edge-hopping
modulation may give rise to the edge state at the perfect armchair
edge.

We can estimate the decay lengths of edge states from a calculation
of the energy levels of a relatively narrow GNR because the
interaction between the two edges of the nanoribbon quickly
decreases as the width of the ribbon increases; the energies of the
edge states then rapidly converge to the semi-infinite limit. The
decay length $\lambda_{\text{decay}} (= 2 \pi/\kappa)$ is obtained
from the decaying factor $e^{-\kappa r_{\perp}}$, where $r_{\perp}$
is the perpendicular direction to the boundary. The estimated decay
lengths are 1.23 and 1.59 nm for states A$'$ and B$'$, respectively,
as obtained from gradients of $-$0.510 and $-$0.394 \AA$^{-1}$, respectively, in
the semilog plot of the decaying electronic densities in
Fig.~\ref{Fig2}. The shorter decay length of state A$'$ can also be
seen in Fig.~\ref{Fig2}(b).
\begin{figure}[t]
\includegraphics[width=0.8\columnwidth]{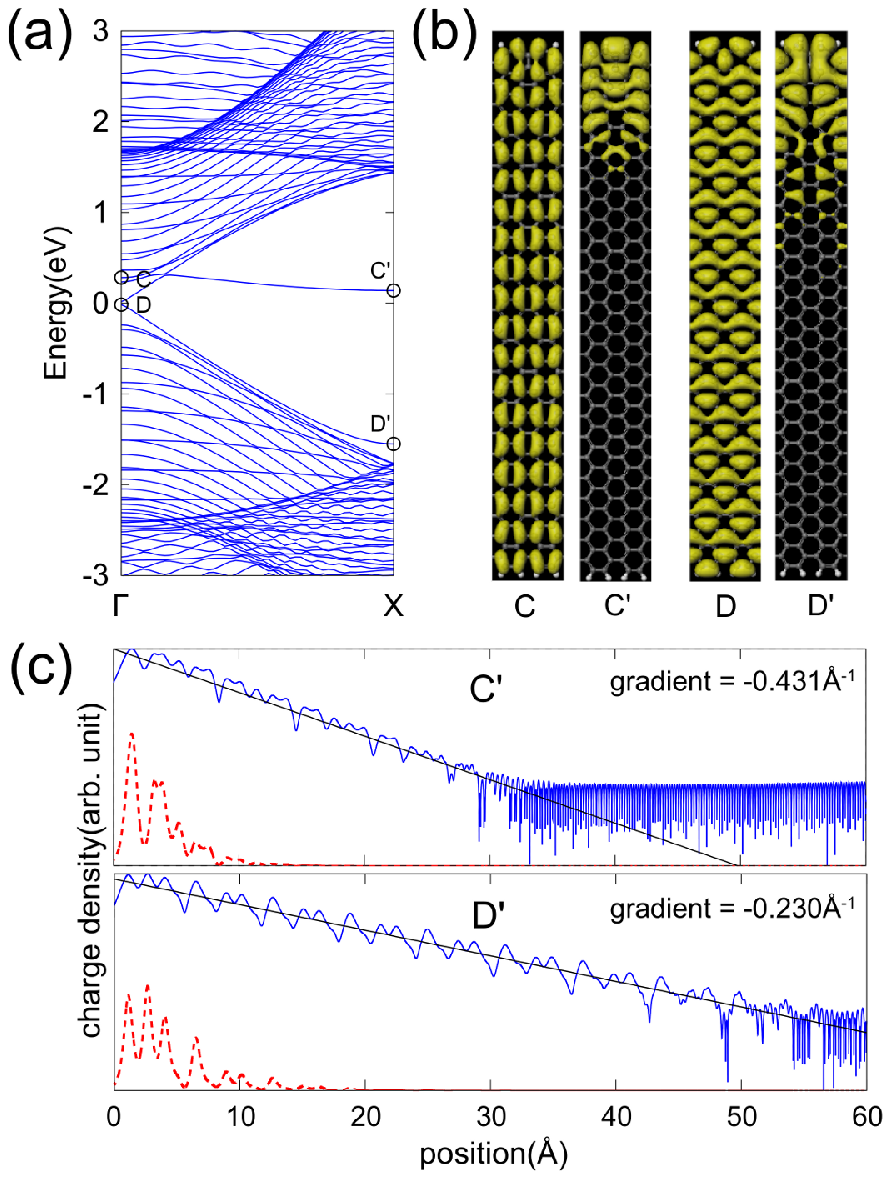}
\caption{(Color online) (a) Band structure of a GNR with the ac(677) edge, and (b)
electronic densities of two edge-related states. The labels C (D)
and C$'$ (D$'$) indicate electronic densities in the same band at
$\Gamma$ and X, respectively. (c) Plane-averaged electron densities
(red) and their semilog plots (blue) for the GNR with the ac(677)
edge showing an exponential decaying character; the gradients of
fitted curved give the decay parameters in units of \AA$^{-1}$. The
position indicated on the horizontal axis is the distance from the
hydrogen atom
at the ac(677) edge to the middle of the ribbon.}
\label{Fig3}
\end{figure}

Figure \ref{Fig3} shows the edge states near the Fermi level at the
ac(677)-reconstructed edge. The reconstruction can be understood as
an array of Stone--Wales defects, as this geometry is obtained by
rotating particular carbon--carbon bonds by $90^{\circ}$. The
Stone--Wales defect in carbon nanotubes is known to be able to form
two quasi-bound states\cite{HChoi} that are characterized as bonding
and anti-bonding states in the rotated carbon dimer\cite{GKim}.
Figure \ref{Fig3}(a) shows that a nearly flat band, labeled as
CC$'$, is unoccupied and may act as an acceptor level. In
Fig.~\ref{Fig3}(b), the electronic charge densities of states C$'$
and D have distinct anti-bonding and bonding characteristics in the
rotated carbon dimer, respectively. Although the energy dispersion
of the CC$'$ band is nearly flat, the decay lengths of the edge
states vary significantly depending on $k_{||}$ in the edge
direction. In fact, near the $\Gamma$ point, the edge state is
located inside an allowed energy band so that it is not a localized
state due to mixing with the extended states. For state C$'$, the
energy level is deep into the bulk energy gap. Consequently, its
electronic charge density is strongly localized at the edge and has
a decay length of 1.46 nm. In contrast, the energy level of state
D$'$ is so close to the valence band edge over the whole $k_{\|}$
range that the decay length is relatively long (2.73 nm). The
difference in the decay lengths can also be seen Figure
\ref{Fig3}(b).

We can deduce the decay length of an edge state from its energy
level when the Bloch wavevector {\bf k} is assumed to be a complex
number. The general procedure for considering the complex wavevector
is as follows: The crystal translational symmetry produces a
wavevector {\bf k} a good quantum number, and the bulk Hamiltonian
can be decoupled for each {\bf{k}}. The secular equation
\begin{eqnarray}
det[E-H({\bf{k}})]=det[E-H(k_\|,k_{\perp})]=0
\end{eqnarray}
then gives the energy levels and Bloch wavefunctions of the crystal.
In this case, {\bf{k}} is regarded as a parameter. If we now also
regard the energy and $k_\|$ as parameters and solve the secular
equation with respect to $k_{\perp}$, then the secular equation
becomes a polynomial of $k_{\perp}$ in general\cite{superlattice}.
In line with the fundamental theorem of algebra, the polynomial has
the same number of zeroes (including complex values) as the degree
of the polynomial independent of the energy and $k_\|$. This means
that whenever there is a band edge where the real band begins to
vanish, a corresponding complex band appears from the band edge. In
view of the single-particle Schr\"odinger equation, the wavefunction
at the boundary should be matched to a linear combination of bulk
wavefunctions (including waves having complex wavevectors). When
this linear combination does not include any wavefunctions with a
purely real wavevector, it forms surface or interface evanescent
states.

\begin{figure}[t]
\includegraphics[width=0.8\columnwidth]{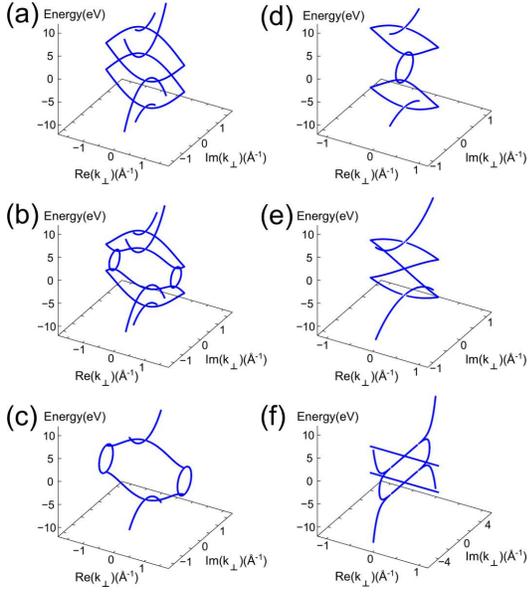}
\caption{(Color online) Tight-binding complex band structures of graphene in the
armchair direction (a) $k_{||} = 0$, (b) $k_{||} =
\frac{1}{4}G_{||}$, and (c) $k_{||} = \frac{1}{2}G_{||}$, where
$G_{\|}=\frac{2\pi}{3a}$, and $a$ is the carbon--carbon bond length
(= 1.42\AA), and the zigzag direction (d) $k_{||} = 0$, (e) $k_{||}
= \frac{1}{3}G_{||}$, and (f) $k_{||} = \frac{1}{2}G_{||}$, where
$G_{\|}=\frac{2\pi}{\sqrt{3}a}$. The hopping energy is set to
$-2.88$ eV. } \label{Fig4}
\end{figure}

Near the band edge, the behavior of the complex band is easily
derived from an effective Hamiltonian. Usually, the real band has an
approximately quadratic dispersion near its edge, and the effective
Hamiltonian can be expressed as
\begin{eqnarray}
H({\bf{k}})=\frac{\hbar^2 |{\bf k}|^2}{2m}+E_0=\frac{\hbar^2}{2m}(k_{\|}^2+k_{\perp}^2)+E_0.
\end{eqnarray}
For a given $k_{\|}=k_0$ and $E$, solving this Hamiltonian with
respect to $k_{\perp}$ gives the complex band structure. When $E$
lies in the forbidden energy regime ($E<\frac{\hbar^2
k_0^2}{2m}+E_0$),
\begin{eqnarray}
k_{\perp}=\pm i \sqrt{k_0^2-\frac{2m(E-E_0)}{\hbar^2}},
\end{eqnarray}
and the decaying parameter $k_{\perp}\propto \sqrt{E_0'-E}$, where
$E_0'\equiv\frac{\hbar^2 k_0^2}{2m}+E_0$ is the band edge for a given $k_{\|}$.

The nearest-neighbor tight-binding model provides the essential
features of the complex band structure and assists in understanding
the overall structures of the complex bands
We can solve the secular equation
of a $4\times 4$ tight-binding matrix for $k_{\perp}$ and obtain
analytic solutions for specific $k_{||}$ and $E$ values.
Within the framework of the tight-binding approximation, we can
obtain the decay parameters for states A$'$, B$'$, C$'$, and D$'$ in
the ac(56) and ac(677) edges (Figs. 2 and 3). At
$k_{||}=\frac{1}{2}G_{\|}=\frac{\pi}{3a}$, where $a$ is the
carbon--carbon bond length, the decay parameters for states A$'$ and
B$'$ are 0.375 and 0.325 \AA$^{-1}$, respectively. On the other hand, the
decay parameters for states C$'$ and D$'$ are 0.297 and 0.196
\AA$^{-1}$, respectively, at
$k_{||}=\frac{1}{4}G_{\|}=\frac{\pi}{6a}$.
Although the nearest-neighbor tight-binding method can provide
overall structures of the complex band, there is a discrepancy
between the decay lengths calculated using the tight-binding and
first-principles methods for each state. Unlike the real band
structure that can accurately describe the energy region near the
Fermi level, the complex band structure derived from the
tight-binding calculation is not accurate because it depends
sensitively on the energy positions of the band edges at each
$k_{\|}$, which is usually far from the Fermi level.

In Fig.~\ref{Fig4}, the tight-binding complex band structures
are plotted for particular $k_{\|}$ values
in both the armchair and zigzag directions. In both cases, the
complex band has a quadratic shape near the X point (the band edge),
the localization becomes stronger further away from the band edge,
and reaches a maximum deep inside the band gap. At any energy around
the Fermi level, there are always four (two) $\pi$-bands in the
graphene in the armchair (zigzag) directions. This feature is
related to the different number of maximum scattering channels in
the two directions because the energy and crystal momentum along the
direction parallel to the edge are conserved during electron
scattering\cite{PNAS}.

\begin{figure}[t]
\includegraphics[width=0.8\columnwidth]{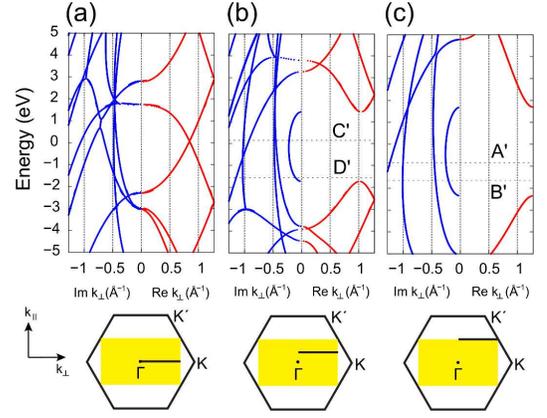}
\caption{(Color online) First-principles complex band structures of graphene in the
armchair direction. Real and imaginary bands are plotted along the
positive and negative axes, respectively. The lower hexagons are the
BZ of graphene, and the yellow rectangles denote the BZ if a doubled
unit cell of graphene is considered for the armchair cell
construction. The band structures are plotted along line segments.
From the left panel, (a) $k_{||} = 0$, (b) $k_{||} =
\frac{1}{4}G_{||}$, and (c) $k_{||} = \frac{1}{2}G_{||}$, where
$G_{\|}=\frac{2\pi}{3a}$.
In (b) and (c), the dotted lines represent the energy levels of the
localized edge states A$'$, B$'$, C$'$, and D$'$ in Figs. \ref{Fig2}
and \ref{Fig3}, respectively.
}
\label{Fig5}
\end{figure}

\begin{figure}[t]
\includegraphics[width=0.9\columnwidth]{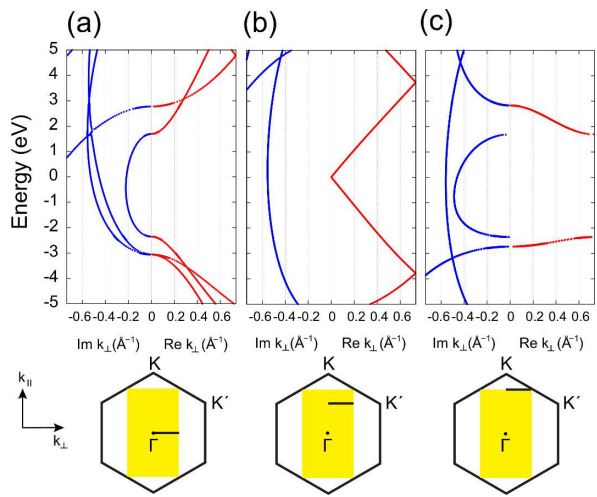}
\caption{(Color online) First-principles complex band structures of graphene in the
zigzag direction. The yellow rectangle denotes the BZ of a doubled
unit cell of graphene for the zigzag rectangular cell construction.
From the left panel, (a) $k_{||} = 0$, (b) $k_{||} =
\frac{1}{3}G_{||}$, and (c) $k_{||} = \frac{1}{2}G_{||}$, where
$G_{\|}=\frac{2\pi}{\sqrt{3}a}$, and $a$ is the carbon--carbon bond
length. } \label{Fig6}
\end{figure}

Complex band structures of graphene in the armchair direction
calculated from first principles in a plane-wave basis are shown in
Fig.~\ref{Fig5} for three $k_{||}$ values ($k_{||} = 0$,
$\frac{1}{4}G_{||}$, and $\frac{1}{2}G_{||}$), where
$G_{\|}=\frac{2\pi}{3a}$. In principle, any edge state in the bulk
region can be represented by a linear combination of wavefunctions
that correspond to a complex band of the same energy, and in
practice, the dominant contribution comes from the first few bands
with long decay lengths. Furthermore, in case of the graphene edge,
the first complex band with the longest decay length is composed of
$\pi$-electrons and determines the decay length of the edge state.
The energy levels of the edge states of the ac(56) edge at the X
point (A$'$ and B$'$ in Fig.~\ref{Fig2}) are $-$0.84 and $-$1.64 eV,
respectively, and the corresponding decay parameters ($\kappa$) are
0.255 and 0.207 \AA$^{-1}$, as shown in Fig.~\ref{Fig5}(c). Because
the charge density is given by the squared magnitude of the
wavefunction, the decay parameters ($\kappa$) should be doubled for
comparison with the gradients in Fig.~\ref{Fig2}: this gives decay
lengths of 0.510 and 0.394 \AA$^{-1}$ for A$'$ and B$'$,
respectively. The small difference is attributed to the finite size
effect of the ribbon width.

In the case of the ac(677) edge, both $k_{\|}=\frac{1}{4}G_{\|}$ and
$k_{\|}=\frac{1}{2}G_{\|}$ should be considered due to the doubling
of the armchair unit cell. The energy levels of the C$'$ and D$'$
states in Fig.~\ref{Fig3} are 0.14 and $-$1.55 eV, respectively, and
the corresponding decay parameters are 0.219 and 0.110 \AA$^{-1}$ at
$k_{\|}=\frac{1}{4}G_{\|}$ and 0.249 and 0.216 \AA$^{-1}$ at
$k_{\|}=\frac{1}{2}G_{\|}$. Because the complex bands at
$k_{\|}=\frac{1}{4}G_{\|}$ have smaller decay parameter, the decay
lengths are determined by those bands. As mentioned earlier, the
ac(677)-type edge has a lattice parameter that is twice that of the
ac(56)-type edge, and the band edge corresponding to the ac(677)
edge is at $k_{\|}=\frac{1}{4}G_{\|}$. If the $\kappa$ values at
$k_{\|}=\frac{1}{4}G_{\|}$ is doubled, then agreement with the
gradients of the charge densities (0.431 \AA$^{-1}$ for C$'$ and
0.230 \AA$^{-1}$ for D$'$) in Fig.~\ref{Fig3}(c) is obtained.
Here, it is necessary to adjust the energy levels of the GNR to
those of graphene. As the GNR becomes wider, the band gap decreases
at ${\bf k}_{||} = 0$ (${\bf k}_{||} = \frac{2\pi}{3\sqrt{3}a}$) in
the armchair (zigzag) ribbon\cite{YWSon1}. For the undoped case, we
can adjust the center of the band gap of the GNR to the Fermi level
of graphene. If the model GNR is sufficiently wide, the error
becomes negligibly small.

We would like to stress here that the relationship between the
complex band and the decay length is applicable to any graphene edge
regardless of its chemical passivation because only the energy
dispersion of the edge state along $k_{\|}$ is needed. In practice,
since the calculated energy quickly approaches the semi-infinite
limit even for a relatively narrow ribbon, we can deduce the decay
length of various graphene edge states from relatively narrow ribbon
calculations.
In the case of the zigzag graphene edge, we must consider the spin
polarization\cite{YWSon2}, and the wavefunctions should also be
matched to the bulk graphene wavefunctions with complex {\bf k}. As
long as the spin density is localized at the edge, the bulk
Hamiltonian for each spin component is almost identical to the
unpolarized spin density so that the decay behaviors of the
spin polarized graphene edge state are also accurately analyzed with
the complex band structures in Figs.~\ref{Fig5} and \ref{Fig6}.

Grain boundaries in polycrystalline graphene, in a similar manner to
reconstructed edges, also have topological defects such as pentagons
and heptagons. At a grain boundary, at least two domains are
matched, and the topological defects give rise to localized
electronic states. In such cases, the decay lengths are affected by
the crystallographic direction of the domain. The complex band
structures in the zigzag direction for $k_{||} = 0,
\frac{1}{3}G_{||}$ and $\frac{1}{2}G_{||}$, where
$G_{\|}=\frac{2\pi}{\sqrt{3}a}$, shown in Fig.~\ref{Fig6}, have
considerably different structures from those in the armchair
direction (Fig.~\ref{Fig5}). This unambiguously shows that the decay
length of a localized state depends on the crystallographic
direction of the domain when two distinguishable domains are
connected and a grain boundary is formed. We expect that the decay
behavior of localized states originating from the topological
defects at the grain boundaries is observable using scanning
tunneling spectroscopy.

\section{Conclusion}
In summary, we have shown that it is possible for localized states
to appear in reconstructed armchair graphene edges using {\em ab
initio} pseudopotential calculations. The edge state in the ac(677)
model decays more rapidly than that in the ac(56) model at the X
point. We have also presented complex band structures of graphene in
the armchair and zigzag directions in both the tight-binding and
first-principles frameworks. The extension of the conventional band
structures to a complex band structure provides information on the
energy-dependent decay lengths of the graphene edge states. By
comparing the shapes of the complex bands in the armchair and zigzag
crystallographic directions, we revealed that the decay behaviors of
the edge state are strongly related to the crystallographic
directions. Our analysis indicates that our theoretical approach to
understanding the edge states through the complex band structure is
quite general and can be applied to any graphene-based structure
with edges or grain boundaries.

\section*{Appendix: Tight-binding model for calculating the complex band structure of graphene}
\begin{figure}[t]
\includegraphics[width=0.9\columnwidth]{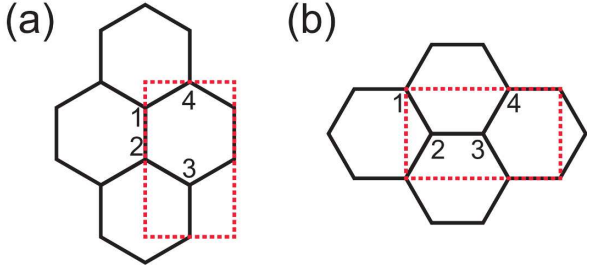}
\caption{(Color online) Doubled unit cells of graphene for the (a) armchair and (b)
zigzag construction. The bases are indexed in the same order as the
$4\times4$ matrix representation of $H(\vec{k})$ in the text.}
\label{Fig7}
\end{figure}

If we construct the unit cell of graphene in the armchair direction
as shown in Fig.~\ref{Fig7}(a), the Hamiltonian can be Fourier
transformed, and each decoupled $H(\vec{k})$ is represented as the
following $4\times 4$ matrix,
$$
\begin{bmatrix}
0 & t & 0 & t(1+e^{-i\vec{k}\cdot \vec{R_x} }) \\
t & 0 & t(1+e^{-i\vec{k}\cdot \vec{R_x} } ) & 0 \\
0& t(1+e^{i\vec{k}\cdot \vec{R_x} } ) & 0 & te^{-i\vec{k}\cdot \vec{R_y} }   \\
t(1+e^{i\vec{k}\cdot \vec{R_x} } ) & 0 & te^{i\vec{k}\cdot \vec{R_y} }&0
\end{bmatrix},
$$
where $\vec{k}=(k_x,k_y), \vec{R_x}=(\sqrt{3}a,0),
\vec{R_y}=(0,3a)$, and $a$ is the carbon--carbon bond length. The
order of the basis ($i$ = 1, 2, 3, and 4) is shown in
Fig.~\ref{Fig7}(a). If we define $r\equiv e^{i\vec{k}\cdot \vec{R_x}
}=e^{i\sqrt{3}Re(k_x)a} \cdot e^{-\sqrt{3}Im(k_x)a} $ and $s\equiv
e^{i\vec{k}\cdot \vec{R_y} }$, the secular equation for a given $E$
and $k_y$ can be written as follows:

\begin{eqnarray}
0&=&\frac{det(E-H)}{t^4} \nonumber \\
&=&
\left(r+\frac{1}{r}\right)^2
+\left\{4-2Re(s)-2\left(\frac{E}{t}\right)^2 \right\} \left(r+\frac{1}{r}\right)   \nonumber \\
&+&5-4Re(s)+\left(\frac{E}{t}\right)^4 - 6\left(\frac{E}{t}\right)^2 .
\end{eqnarray}

\noindent Because $r\ne 0$, this equation is a fourth-order
polynomial of $r$ and has four solutions. Equation (5) is invariant
under the operations $r \rightarrow \frac{1}{r}$ and $\frac{1}{r}
\rightarrow r$. If $r_s$ is a solution to Eq. (5), then $r_s^*,
\frac{1}{r_s}$, and $\frac{1}{r_s^*}$ are also solutions of the
secular equation. If the solution has a nonzero $\kappa$ ($= Im \
k_x$), then the four solutions correspond to left- and
right-decaying states and their time-reversal pairs. The existence
condition of the complex band can be derived if $\kappa$ is set to
zero.

To visualize the energy-dependent decay length, consider a special
case of the band edge: $k_y=\frac{1}{2}G_{\|}=\frac{\pi}{3a}$. In
this case, if we put $s=-1$, the equation becomes
\begin{equation}
\left\{r+\frac{1}{r}+3-\left(\frac{E}{t}\right)^2 \right\}^2=0.
\end{equation}
The solution of equation (6) is
$$
\sqrt{3}k_xa=
\begin{cases}
\pm\pi \pm i\cosh^{-1}\left\{\frac{3}{2}-\frac{1}{2}\left(\frac{E}{t}\right)^2\right\},
& E^2\le t^2  \\
\pm\cos^{-1}\left\{-\frac{3}{2}+\frac{1}{2}\left(\frac{E}{t}\right)^2\right\},
& t^2<E^2\le 5t^2    \\
\pm i\cosh^{-1}\left\{-\frac{3}{2}+\frac{1}{2}\left(\frac{E}{t}\right)^2\right\},
& 5t^2<E^2
\end{cases}
$$
and numerical solutions are plotted for the case of $k_y=0,
\frac{\pi}{6a}$, and $\frac{\pi}{3a}$ in Fig.~\ref{Fig4}(a), (b),
and (c), respectively.

In the same manner, for graphene in the zigzag direction, the
Hamiltonian $H(k_x,k_y)$ is given by
$$
\begin{bmatrix}
0 & t(1+e^{i\vec{k}\cdot \vec{R_y} }) & 0  & te^{-i\vec{k}\cdot \vec{R_x} } \\
t(1+e^{-i\vec{k}\cdot \vec{R_y} }) & 0 & t & 0 \\
0& t & 0 & t(1+e^{-i\vec{k}\cdot \vec{R_y} })   \\
te^{i\vec{k}\cdot \vec{R_x} } & 0 & t(1+e^{i\vec{k}\cdot \vec{R_y} }) &0
\end{bmatrix},
$$
where $\vec{k}=(k_x,k_y)$, $\vec{R_x}=(3a,0)$, and
$\vec{R_y}=(0,\sqrt{3}a)$. The order of the basis ($i$ = 1, 2, 3,
and 4) is shown in Fig.~\ref{Fig7}(b). If we define $r\equiv
e^{i\vec{k}\cdot \vec{R_x} }$ and $s\equiv e^{i\vec{k}\cdot
\vec{R_y} }$, then secular equation for a given $E$ and $k_y$ can be
written as follows:
\begin{eqnarray}
0&=&\frac{det(E-H)}{t^4} \nonumber \\
&=&
5-\{2+2Re(s)\}\left(r+\frac{1}{r}\right)
+4\{Re(s)\}^2 \nonumber \\
&+&\left\{8-4\left( \frac{E}{t} \right)^2 \right\}Re(s)
+\left(\frac{E}{t}\right)^4 - 6\left(\frac{E}{t}\right)^2 .
\end{eqnarray}
With respect to the variable $r$, this equation has two solutions connected by an inverse relation.
From this equation we can calculate the decay length of the edge
states of the GNR. Because the edge state of a zigzag-edged GNR is a
zero-energy mode, the $k_y$-dependence of the decay length is
obtained by solving the following equation,
\begin{equation}
\{2+2Re(s)\}\left(r+\frac{1}{r}\right) =\{2+2Re(s)\}^2+1.
\end{equation}
The solution of equation (8) is $k_xa=\pm \frac{i}{3}\ln
\left\{4\cos^2\left(\frac{\sqrt{3}k_ya}{2}\right)\right\}$, and
numerical solutions are plotted for the cases of $k_y=0,
\frac{2\pi}{3\sqrt{3}a}$, and $\frac{\pi}{\sqrt{3}a}$ in
Fig.~\ref{Fig4}(d), (e) and (f), respectively.

\section*{Acknowledgments}
G.K. acknowledge the support of the Basic Science Research Program through the National Research Foundation of Korea (NRF) funded by the Ministry of Education
(Grant No. 2013R1A1A2009131) and the Priority Research Center Program (Grant No. 2010-0020207) of the Korean Government.
C.P. and J.I. were supported by NRF (Grant No. 2006-0093853).
Computations were performed through the support of KISTI.

\end{document}